\documentclass[journal,comsoc]{IEEEtran}
\usepackage[T1]{fontenc}
\usepackage{cite}
\usepackage{graphicx}
\usepackage[english]{babel}
\usepackage[font={small}]{caption}
\usepackage{color}
\usepackage{newfloat}
\DeclareFloatingEnvironment[fileext=lop]{Table}
\begin{document}
\makeatletter

\makeatother
\title{Toward RIS-Enhanced Integrated Terrestrial/Non-Terrestrial Connectivity in 6G}

\author{Parisa Ramezani, Bin Lyu, and Abbas Jamalipour}

\maketitle

\begin{abstract}  

The next generation of wireless systems will take the concept of communications and networking to another level through the seamless integration of terrestrial, aerial, satellite, maritime and underwater communication systems.  Reconfigurable intelligent surface (RIS) is an innovative technology which, with its singular features and functionalities, can expedite the realization of this everywhere connectivity. Motivated by the unparalleled properties of this innovatory technology, this article provides a comprehensive discussion on how RIS can contribute to the actualization and proper functioning of future integrated terrestrial/non-terrestrial (INTENT) networks. As a case study, we explore the integration of RIS into non-orthogonal multiple access (NOMA)-based satellite communication networks and demonstrate the performance enhancement achieved by the inclusion of RIS via numerical simulations.  Promising directions for future research in this area are set forth at the end of this article.

\end{abstract}

\IEEEpeerreviewmaketitle

\section{Introduction}
As the deployment of the fifth-generation (5G) networks continues in different parts of the world, the research community has started focusing on the sixth-generation (6G) systems and how this generation can revamp wireless networks.  One of the most important vows of 6G is to provide everywhere connectivity for anyone and anything through the effectual integration of terrestrial and non-terrestrial networks \cite{Vaezi}. This ubiquitous connectivity must be accompanied with high spectral efficiency (SE) and energy efficiency (EE), ultra-low latency, large capacity, high security, etc. to support the resource-intensive applications of 6G and offer an excellent level of quality of experience (QoE) for heterogeneous 6G users \cite{Kunst}. The path to this grandiose goal is not free of challenges and major breakthroughs are urgently called for.  

Very recently, the groundbreaking technology of reconfigurable intelligent surface (RIS) has sprung up which, with its unique features, is expected to be a major actor in the evolution of the next generation systems. RIS is a software-controlled metasurface, composed of semi-passive elements which can be dynamically modified to shape the propagation environment.  The far-reaching effects of RIS on the behavior of wireless medium can make this technology an imperative constituent of the upcoming 6G-enabled \textit{integrated terrestrial/non-terrestrial} (INTENT) environments, where seamless everywhere connectivity is demanded with a high performance and at a reasonable cost.

The advantages of RIS over multi-antenna techniques and relaying methods have been thoroughly discussed in the literature for terrestrial communication systems. These advantages include low energy consumption, high SE, reduced complexity, low production cost, etc. The same arguments are valid for non-terrestrial networks as well. Precisely, unmanned aerial vehicles (UAVs), high-altitude platforms (HAPs), and satellites have constraints on weight, size, and power consumption. Therefore, employing massive multiple-input multiple-output (MIMO) technologies and active relaying techniques does not seem to be efficient. RIS, with its simple hardware, light weight, full-duplex (FD) functionality, and low energy consumption is a very good candidate for assisting these platforms and extending their operational lifetime. Besides, for underwater communication, it is very important to devise solutions with low energy consumption and complexity because of the limited recharging and energy harvesting opportunities in the underwater environment. Therefore, RIS is envisaged to be a game-changer for improving the performance of underwater communications \cite{Kisseleff}.

Given the high potentials of RIS to efficiently connect different segments of the ground, air, space, and water, it is imperative to have a closer look into the role of RIS in the imminent 6G-enabled INTENT ecosystem. Motivated by this, the aim of this article is to answer the following three important questions:

\begin{enumerate}
\item \emph{How can RIS be implemented in different segments of the INTENT environment?}

\item \emph{In what ways can RIS contribute to the advancement of 6G-enabled INTENT networks?}

\item \emph{What are the main challenges in the way of achieving a successful integration of RIS and INTENT networks?}
\end{enumerate}

To answer the above questions, we present the detailed architecture of the prospective RIS-enhanced INTENT networks, explicate the direct and indirect roles of RIS for the development of INTENT systems, and delineate the challenges that must be overcome for the smooth functioning of RIS in all dimensions of this integrated ecosystem. We also present a case study where the possible gain of using RIS for assisting an uplink satellite communication system is investigated.

\begin{figure*}[t!]
\centering
\includegraphics[width=7.4in]{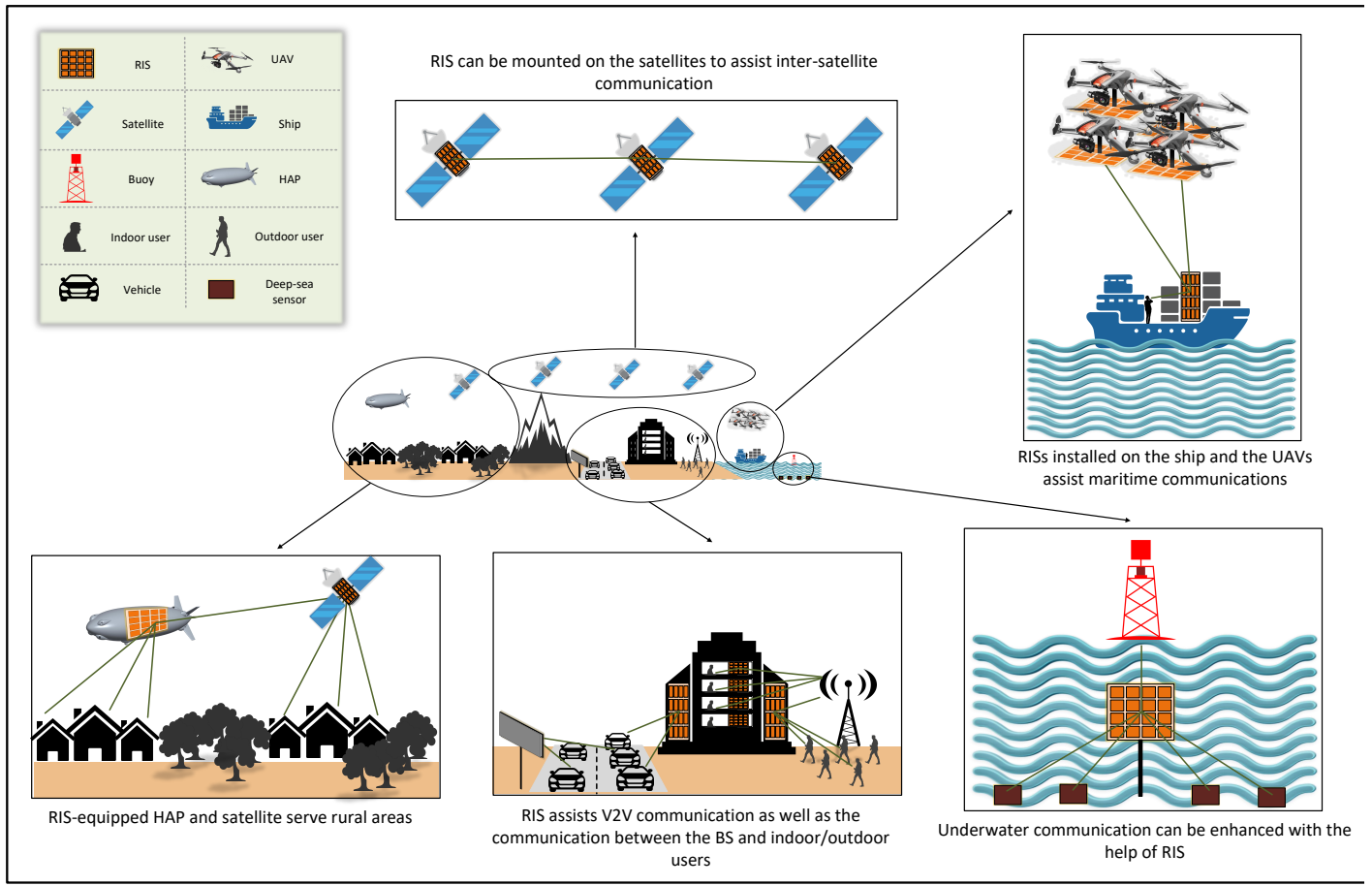}
\caption{ Architecture of RIS-enhanced INTENT networks.}
	\label{arch}
	
\end{figure*}

The main contributions of this study are summarized as follows:
\begin{itemize}

\item We introduce the architecture of the INTENT environment enhanced by RIS and provide discussions on the implementation and feasibility of RIS-aided networks in different dimensions. 

\item We present a comprehensive discussion on the direct role of RIS for the all-around performance improvement of 6G-enabled INTENT networks.
  
\item We elaborate on how RIS can contribute to the progress of INTENT networks by improving the performance of wireless powered communication (WPC),  non-orthogonal multiple-access (NOMA), backscatter communication (BackCom), and terahertz communication (THzCom) technologies. 

\item A novel case study is provided for RIS-aided NOMA-based satellite communication to highlight the gain that RIS brings into this type of communication. 

\item Insightful future research guidelines are provided by introducing seven fundamental research challenges. 
\end{itemize} 

The rest of this article is structured as follows: First, the architecture of the future RIS-enhanced INTENT networks is introduced, followed by a detailed discussion on the direct and indirect roles of RIS for the advancement of these networks. A case study is presented to illustrate the potential performance enhancement brought by RIS. Then, some fundamental research directions are specified. Finally, concluding remarks are provided. 

\section{Architecture of RIS-Enhanced INTENT Networks }
Figure \ref{arch} illustrates an example architecture for the impending RIS-enhanced INTENT networks, where RIS is deployed and utilized in different parts of the integrated environment in order to enhance the network performance in all dimensions. 
 
As shown in Figure \ref{arch}, building facades and windows can be coated with RIS to serve outdoor and indoor users, respectively. In addition, the RISs mounted on outdoor structures such as billboards and building walls assist in vehicle-to-vehicle (V2V) communications by establishing reliable links between the vehicles. 

RIS-equipped UAVs passively reflect the signals towards the intended receivers without needing to consume excessive energy. These UAVs are of specific importance in cases where direct communication is hard to establish, e.g., maritime communication between the ship and terrestrial base station (BS). The ship-borne RISs can further enhance the performance of maritime communications and offer an improved QoE for traveling users. Additionally, underwater acoustic communication suffers from severe path-loss and frequency-selectivity due to multipath propagation. Based on a recent study, utilization of RIS is beneficial for alleviating the multipath effect in the underwater medium, where the orientation of RIS elements can be optimized to mitigate the frequency-selectivity and improve the effective signal bandwidth \cite{Kisseleff}.

In satellite communication, passive reflect-arrays with fixed phase shift configurations have been traditionally used for combatting the extreme path-loss due to very large communication distances. These reflect-arrays can be substituted by RIS in the forthcoming 6G-enabled satellite communication systems to enable dynamic phase adjustment \cite{Zheng}. Moreover, RIS-aided inter-satellite communication is a potent strategy for improving the performance of satellite communication networks  in next-generation systems \cite{Tekbiyik1}. Similarly, HAPs, which are typically used for providing connectivity to remote areas, can be enhanced by RIS in order to transmit or relay signals with improved directivity and less power consumption. 

\subsection{Implementation}

RIS can coat the surface of a HAP and be installed as a horizontal surface at the bottom of the UAV \cite{Alfattani,Shang}. For deployment on the satellites, the large area under the solar panels can be leveraged for RIS installation \cite{Tekbiyik2}. RIS deployment in the underwater medium can be realized in various ways; for example, RIS can be attached to the ground or float beneath the water \cite{Kisseleff}. Further, the walls and windows of the ships as well as the outer surface of autonomous underwater vehicles (AUVs) can be coated with RIS.

The phase shifts of terrestrial RIS are calculated by the terrestrial BS and fed back to the elements by a smart controller. Analogously, for the RISs deployed on the ship, the optimal phase shifts are calculated by the ship station; the RIS controller then applies the reconfiguration. For the RISs mounted on the UAVs, HAPs, and satellites, the phase shift calculations are performed by the onboard processing units of the platforms. Then, the reconfiguration of the UAV- and HAP-borne RIS is conducted by the RIS central unit which is a part of the onboard aerial platform's controller \cite{Alfattani}, while the transceiver of the satellite takes over the role of the controller for the installed RIS \cite{Zheng}. As for the RISs deployed underwater, buoys with high processing capabilities do the necessary calculations for finding the optimal RIS phase shifts. These buoys also host the network controller for controlling the RIS and other parts of the underwater communication network \cite{Akyildiz}.

\subsection{Are RIS-Enhanced INTENT Networks Feasible?}

RIS is a thin surface of lightweight elements and can be designed in arbitrary shapes depending on the intended deployment place. It consumes ultra-low power, possesses simple hardware, and has low production cost. Therefore, RIS accords well with the size, weight, power, and cost requirements of future 6G networks and offers a high level of flexibility for deployment in all dimensions of the INTENT ecosystem. Thus, the short answer to the question in title is YES. However, a number of issues have to be addressed before the wide-scale adoption of RIS-assisted networks. While we will comprehensively discuss the relevant challenges of RIS-enhanced INTENT networks in a later section, it is worth a short discussion here. 

The feasibility and proper functioning of RIS-assisted INTENT networks highly depends on the accuracy of the assumptions made for studying the performance of these networks. For instance, perfect channel state information (CSI) availability is a frequent assumption in the RIS literature which must be carefully revisited for studying RIS-enhanced INTENT networks. Real-time reconfigurability of RIS elements is another common assumption, the feasibility of which is yet to be verified via practical testbed experiments. Taking the RIS-aided underwater communication as an example, novel optimization methods and algorithms are required to update the optimal RIS configuration in very short time intervals so as to adapt to the small coherence time of the underwater medium. Hence, the long answer to the above question is that the vision of RIS-enhanced INTENT networks can be realized if the existing challenges are attentively addressed. At the end of this article, we will provide details on critical issues which promptly call for effective solutions. 

\begin{figure*}[t!]
\centering
\includegraphics[width=7.4in]{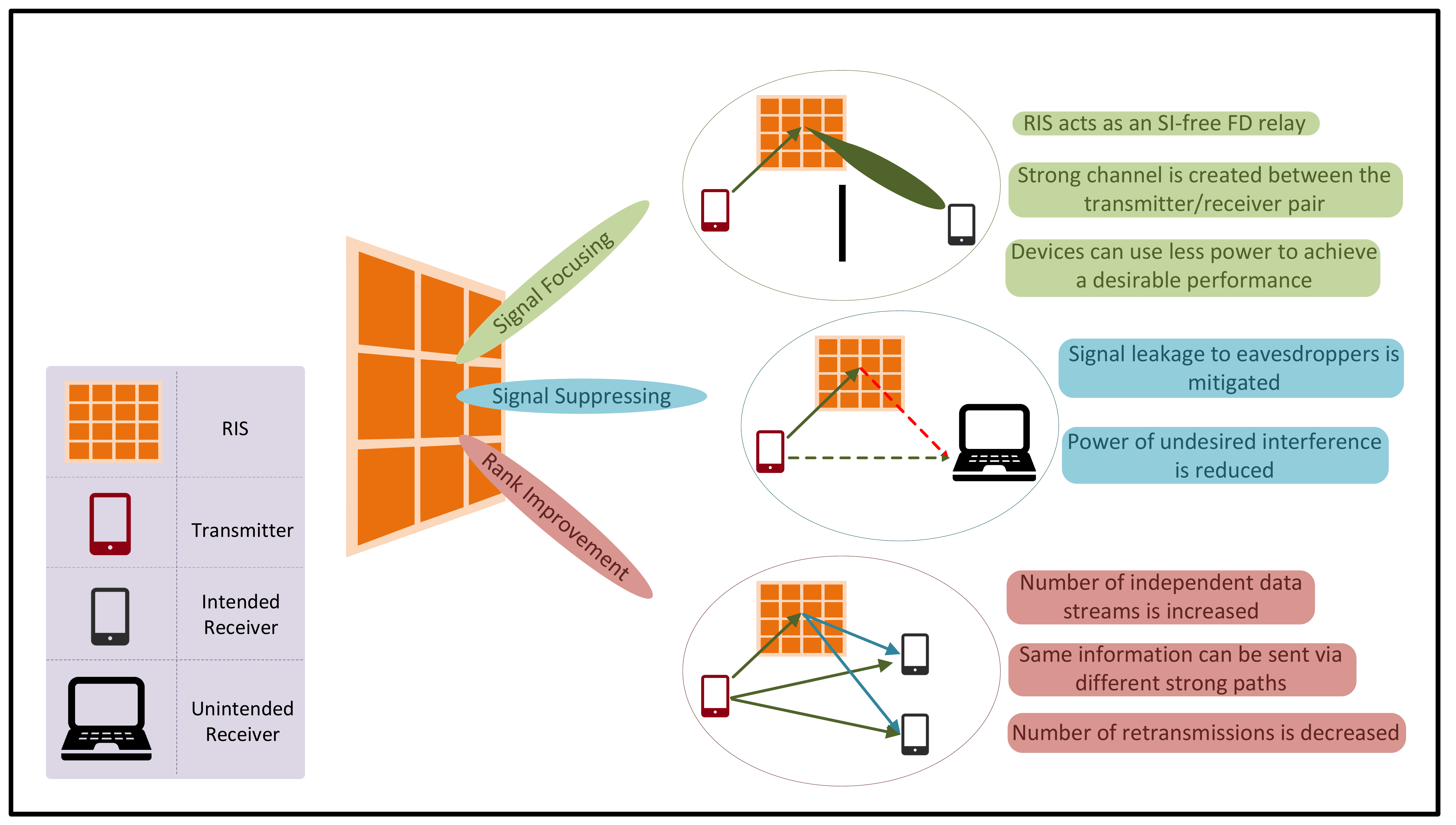}
\caption{ The role of RIS for improving the communication performance of 6G-enabled INTENT networks.}
	\label{ris-func}
	
\end{figure*}

\section{The Direct Role of RIS in 6G-Enabled INTENT Networks}
RIS has many-sided benefits for advancing 6G-enabled INTNENT systems. It can act as an auxiliary network element for boosting the communication performance, perform sensing tasks through its elements, and take the role of a massive MIMO transmitter.  

\subsection{RIS as a Communication Helper}
\subsubsection{Signal Focusing}
RIS elements can modify the signals incident on the surface in a unified manner, such that the signals reflected by RIS elements are constructively combined with those of the direct path. This RIS functionality can introduce multifaceted performance improvements into INTENT networks. SE enhancement is one of the most significant gains that can be attained thanks to the RIS's ability to boost the power of the received signals. Coverage extension is another important benefit that can be brought by RIS into INTENT networks. For instance, in maritime communications, when the ship gets farther from the shore, its connection with the onshore BS gradually degrades. UAVs can be used as active relays to  facilitate the communication between the ships and land-based BSs; however, a more efficient strategy would be to use RIS-equipped UAVs which essentially work in an FD manner without needing complex self-interference (SI) cancellation techniques or suffering from residual SI. Moreover, RIS can very well comply with the UAV constraints; this is while active relays incur excessive burden on the UAVs as they require complex circuits, power amplifiers, etc.  Maintaining connectivity in underwater communications is also very challenging because acoustic signal propagation in the underwater media suffers from substantial signal scattering and absorption loss. The inclusion of RIS has been lately suggested for tackling the impaired connectivity in underwater communications\cite{Kisseleff}. Additionally, RIS can be of remarkable importance for reducing the end-to-end delay of communication. Taking intelligent transportation system (ITS) as an example, real-time information must be delivered on time in order to ensure efficient operation of vehicular applications like autonomous driving. However, the communication in vehicular environments is usually susceptible to frequent link interruptions which make direct vehicle-to-everything (V2X) communications challenging or even impossible. The proper deployment of RIS can noticeably reduce the communication delay and help in timely delivery of delay-sensitive information. Furthermore, RIS can bring significant EE gains into future INTENT networks by letting communicating devices achieve a desirable performance without consuming large amounts of energy. As an example, the utilization of RIS can be very helpful for reducing the energy consumption of the sensors deployed in harsh environments such as underwater\cite{Kisseleff}.

\subsubsection{Signal Suppressing}
One of the most important functionalities of RIS is its ability to suppress signals at unintended receivers by which RIS elements collaboratively modify the incident signals such that the reflected signals cancel out the direct ones. This way, RIS can help suppress information leakage to illegitimate receivers. In satellite communications, for example, the downlink transmission from the satellite to ground users is prone to eavesdropping threats which can be alleviated with proper deployment of RIS. Moreover, by suppressing the undesired interference, RIS remarkably improves the signal-to-interference-plus-noise ratio (SINR) and boosts SE, especially in interference-limited communication systems. 

\subsubsection{Rank Improvement}
 RIS is able to improve the rank of the channel by creating additional distinctive paths, thereby, substantially improving the multiplexing gain and increasing the system capacity. This is very important for facilitating the massive connectivity in Internet of Everything (IoE) networks, where a large number of indoor and outdoor users need to be concurrently served. Further, by adding strong propagation paths between the transmitter and the receiver, RIS increases the diversity gain and elevates the chance of successful information transmission, which enhances the reliability of communication. What is more, as RIS improves the communication reliability, the success rate of information transmission is increased which in turn reduces the required packet retransmissions and results in lower end-to-end transmission latencies. This is favorable for delay-intolerant applications such as online gaming, which are promised to be supported for all users including those aboard airplanes and ships. The improved reliability that RIS renders to INTENT networks also leads to EE enhancement as less energy needs to be expended on packet retransmissions. 

A schematic overview of RIS functionalities and their implications for enhancing the performance of INTENT networks is provided in Figure \ref{ris-func}. Furthermore, Table \ref{tab} specifies the most relevant RIS features for improving the network performance from the view of different metrics.
\begin{Table}[t!]
\centering
\includegraphics[width=3.4in]{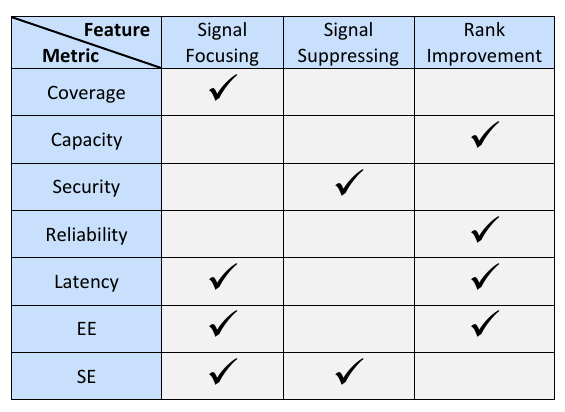}
\caption{Communication performance metrics and related RIS features.}\label{tab}
	\end{Table}

\subsection{RIS as a Sensing Platform}

RIS can be used as a sensing platform for monitoring the environment through its elements. By piggybacking the sensed data into the signals impinging on the surface, RIS forms a symbiotic radio system and presents an energy- and spectrum-efficient alternative to conventional sensors which need to consume extra power and bandwidth for reporting their data. For instance, a UAV-mounted RIS intended for assisting the communication between two ground entities can be also involved in sensing tasks with its elements monitoring the air pollution index. This way, the limited resources of the UAV can be optimally utilized for two purposes without encumbering the UAV with any extra active relaying and sensing equipment. 

\subsection{RIS as a Virtual Massive MIMO Transmitter}
Being illuminated by a nearby radio frequency (RF) source, RIS can modulate data onto the unmodulated RF signal and transmit the encoded signal towards the receiver(s), thus forming a virtual massive multi-antenna transmitter while using only one single RF chain \cite{Basar}. A useful example is satellite communication systems where reflect-arrays with fixed phase shifts are normally used for directional transmission, which due to their inflexible configuration, suffer from beam misalignment. Replacing RIS with traditional reflect-arrays, the SE and EE performance of satellite communication can be remarkably improved through the highly dynamic and directive beamforming of the RIS \cite{Zheng,Tekbiyik2}. In addition, the satellite-mounted RIS can modify the properties (e.g., polarization \cite{Yang}) of the illuminating wave such that multiple data streams can be concurrently transmitted by the satellite. 

\begin{figure*}[t!]
\centering
\includegraphics[width=7.4in]{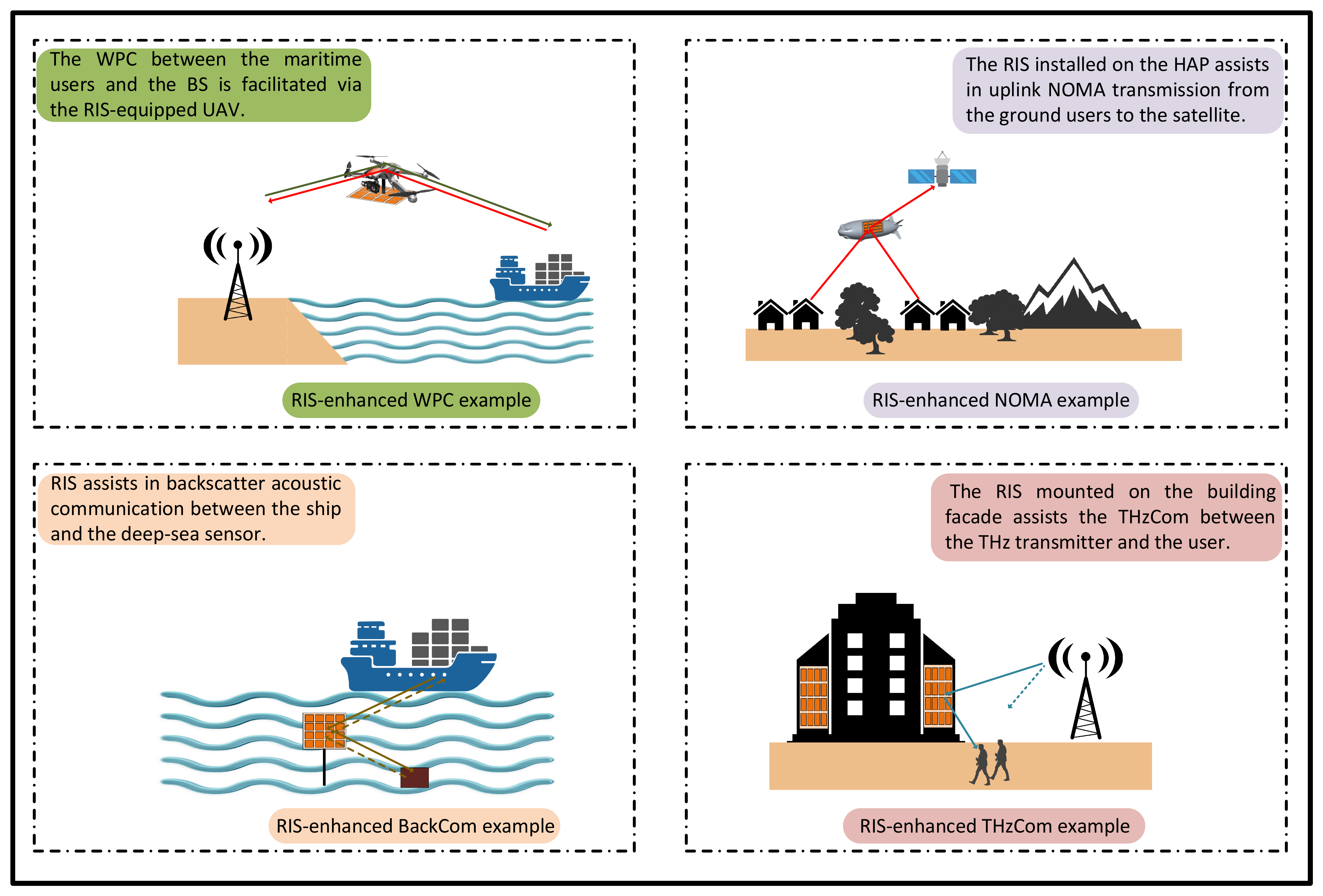}
\caption{ Application of RIS to WPC, NOMA, BackCom, and THzCom systems.}
	\label{ris-ind}
\end{figure*}
\section{The Indirect Role of RIS in 6G-Enabled INTENT Networks }
Besides directly improving the performance of INTENT networks, RIS can take the role of a technology enabler by helping other technologies grow and prosper. There are many promising technologies whose inherent shortcomings may prevent or delay their wide-scale actualization. The integration of RIS with these technologies provides them with a unique opportunity to fulfill their promises and find their place in the upcoming 6G ecosystem. Some of these technologies include WPC, NOMA, BackCom, and THzCom which, as illustrated in Figure \ref{ris-ind}, can be upgraded thanks to RIS. 

\subsection{WPC}
WPC is the technique which allows devices to collect energy from ambient and/or dedicated signals and use the harvested energy to power their communication. This technique can expand the operational lifetime of devices and eliminate the need for human intervention in the process of battery recharging. Underwater and UAV communications are two example scenarios where the use of WPC seems very promising. Specifically, the mission time of underwater nodes and UAVs can be extended via acoustic- and RF-based wireless power transfer (WPT), respectively, without causing any disruptions to their operations. Nevertheless, the low efficiency of WPT has thus far impeded its widespread implementation. In this respect, RIS-empowered WPT is an effective integration of RIS and WPT which can greatly boost the performance of WPC networks.
 
\subsection{NOMA}
NOMA is a communication paradigm which allows the simultaneous transmission from/to a number of users in the same resource block and applies advanced techniques for resolving the interference at the corresponding receivers. Given its potential to enhance the SE and resource utilization, NOMA can be one of the key technologies to serve the ever-growing number of users in all network dimensions. In satellite communication, for example, conventional OMA-based techniques, which are traditionally used to avoid the intra-beam interference among the users, result in low utilization of resources; NOMA can resolve this issue by letting simultaneous transmission to the users. RIS, with its channel reconfiguration capability, can play an important role in the success of the NOMA technology in future INTENT networks \cite{Lyu} because the performance of NOMA largely depends on the channel properties.

\subsection{BackCom}
In BackCom, devices are enabled to communicate with their receivers by modulating and reflecting the incident waves. Owing to its low complexity and cost, this technology is very appealing for the communication of constrained sensors which are being increasingly deployed in all dimensions of the ground, air, space, and water. BackCom has been originally based on modulating and reflecting RF signals; however, it has been recently applied to underwater communication where an acoustic projector transmits acoustic signals letting underwater sensors communicate via modulating and reflecting the acoustic waves \cite{Jang}. The low data-rate of BackCom due to the weakness of the backscattered signal is a serious issue which can hamper the utilization of this innovative technology in the forthcoming 6G-enabled INTENT networks. RIS can come to assist BackCom by strengthening the received signals at backscatter receivers, thereby significantly improving the data-rate \cite{Ramezani}.

\subsection{THzCom}
With the dream of massive ubiquitous connectivity on its verge to become a reality, researchers have recently started to look into the possibility of utilizing the abundant bandwidth in the THz band for communication. However, this frequency band suffers from high propagation loss, penetration loss, and molecular absorption which limit the transmission range. In addition to terrestrial communication, using THz frequencies has been lately proposed for the communication between satellites in space. While molecular absorption is absent in satellite communication systems, misalignment fading due to the sharp beams of THz antennas is a serious problem \cite{Tekbiyik1}. RIS-assisted THz communication is a promising future paradigm, where the aforesaid issues of communicating in the THz band can be coped with by exploiting RIS.

\section{Case Study}
\begin{figure}[h]
\centering
\includegraphics[width=3.4in]{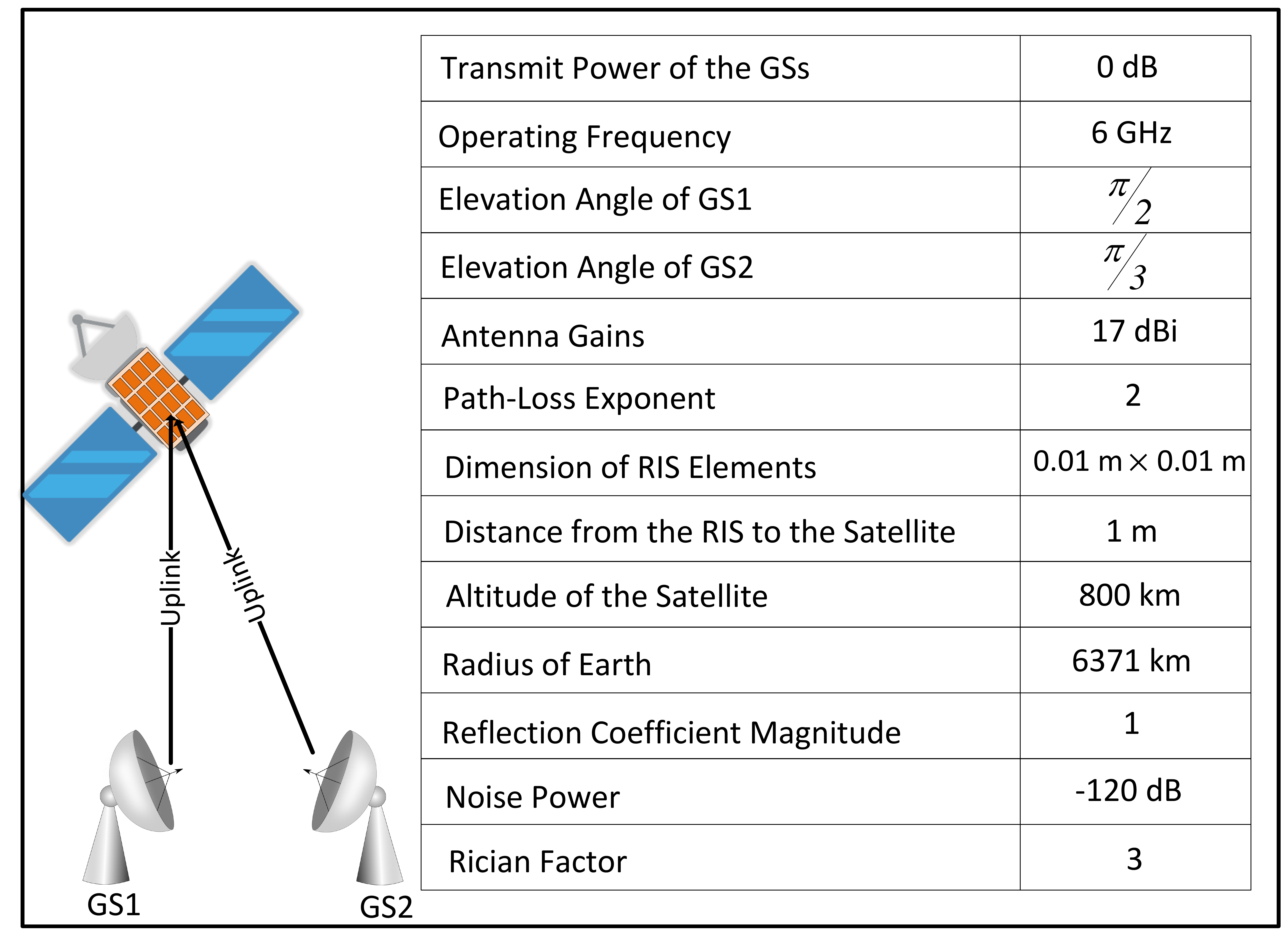}
\caption{ Simulation setup and parameters.}
	\label{sim-set}
	
\end{figure}

\begin{figure}[t!]
\centering
\includegraphics[width=3.4in]{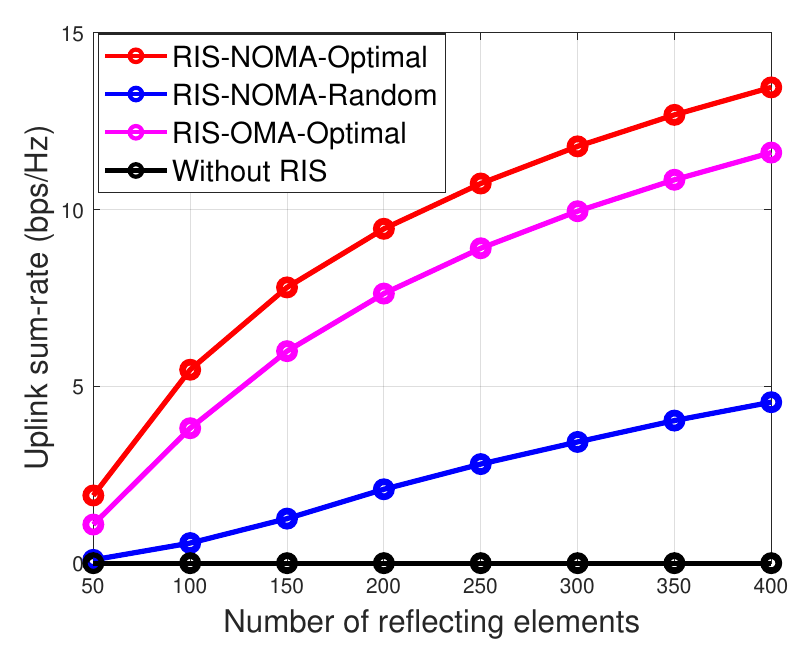}
\caption{ Uplink sum-rate for a satellite communication system with two GSs.}
	\label{sim-res}
	
\end{figure}

As a case study,  a satellite communication system is considered, where two ground stations (GSs) utilize the NOMA technique to simultaneously transmit their information to a satellite which is equipped with an RIS. This resembles the scenario with multiple GSs which are grouped into two clusters, where one GS from each cluster is selected to transmit to the satellite in each transmission sub-block. The path-loss model is taken from reference \cite{Tekbiyik2} and the small-scale fading is Rician-distributed. Simulation setup and parameters are provided in Figure \ref{sim-set}. We have ignored rain attenuation in the simulations. Figure \ref{sim-res} shows the uplink sum-rate performance of the considered RIS-assisted NOMA-based scheme with optimized phase shifts as a function of the number of RIS elements for C-band, where the operating frequency is chosen to be $6$ GHz. Three benchmarks are considered for performance comparison: 1) RIS-assisted NOMA-based scheme with random phase shifts, 2) RIS-assisted orthogonal multiple-access (OMA) scheme, and 3) NOMA-based scheme without RIS. The remarkable advantage of the optimized RIS-based scheme over the schemes with random phase shifts and without RIS clearly demonstrates the important role that RIS is expected to play in the upcoming 6G-enabled satellite communication systems. Furthermore, it can be observed that the NOMA-based approach is superior to its OMA-based counterpart thanks to a more efficient resource utilization. 

\section{Relevant Research Challenges}
Although RIS is foreseen to be a decisive actor for enhancing the performance of the upcoming 6G-enabled INTENT environments, there exist challenges which need to be competently addressed. These challenges include but are not limited to:

\subsection{Channel Modeling} 
One fundamental step towards the successful realization of RIS-enhanced INTENT networks is to propose realistic channel models for performance evaluation and optimization. Different dimensions of the INTENT environments have their own unique characteristics which must be accounted for when developing channel models. For instance, the propagation environment in maritime communications is highly dynamic and prone to atmospheric conditions, whereas the influence of the earth curvature is another factor affecting the maritime propagation medium. As for satellite communications, dynamicity in weather conditions, large Doppler frequency shift and Doppler spread, and long communication distance must be scrutinized when developing channel models. Besides, multipath propagation and high frequency-selectivity are key characteristics of the underwater medium \cite{Kisseleff}, while random UAV fluctuations have a major impact on the quality of the UAV-related channels. Additionally, the channel modeling in RIS-aided systems is still far from mature. Therefore, one promising research direction is to develop practical and scalable channel models with which we can better analyze the performance of RIS-enhanced INTENT networks.

\subsection{Channel Estimation}
The accuracy of CSI is of paramount importance for the reconfiguration of RIS elements. Since the advent of the RIS technology, channel estimation has always been one of the most important challenges in the design of RIS-aided systems. This challenge would be even more serious in case of non-terrestrial networks due to the mobility of aerial and space platforms as well as the fast time-varying channels in maritime and underwater propagation media. So far, only little work has been conducted in the area of channel estimation for RIS-enhanced INTENT networks. \cite{Zheng} is an example which proposes a novel channel estimation framework for RIS-aided satellite communication systems based on the local CSI obtained at the ground users and the satellite. Channel acquisition in RIS-enhanced INTENT networks is thus a major challenge which demands earnest attention from the research community.

\subsection{Effect of Environmental Factors on RIS}
The effect of environmental factors such as temperature variations, rain, and wind on the operation of RIS elements must be thoroughly assessed because improper functioning of some of the RIS elements due to aforementioned factors can bring about serious concerns. For example, cold temperature adversely affects the lifetime of the RIS-embedded batteries which complicates the optimal design of the system since a compromise must be made between performance maximization and energy consumption. In such scenarios, traditional optimization techniques may be inefficient and employment of artificial intelligence-based methods can better cater to the system needs.  

\subsection{Implementation Complexity and Cost}
Deploying RIS in all dimensions of future INTENT networks may result in prohibitive complexity and cost in implementation. Specifically, the inclusion of RIS in different parts of the ground, air, space, and water necessitates a satisfactory level of coordination among all network dimensions. Furthermore, the benefits of RIS for enhancing the performance of INTENT networks very much depends on the ability of RIS elements to be reconfigured in real-time. All of this must be achieved at a reasonable complexity and cost; otherwise, RIS would bring little or no gains as compared to other advanced technologies such as massive MIMO. In order to have an efficient and scalable RIS-aided INTENT environment, resorting to conventional centralized solutions is not logical and designing distributed and lightweight algorithms is imperative.

\subsection{Power Consumption at the RIS}
Though zero power consumption is usually assumed for the RIS, the elements expend non-negligible power for adjusting their reflection parameters. Consequently, the number of active RIS elements for assisting network operations must be meticulously decided to achieve a satisfactory trade-off between RIS power consumption and network performance. The decision on the optimum number of reflecting elements is highly correlated with the installation place of RIS. For instance, if deployed in remote areas with limited fixed BSs,  RIS must be vigilant in its energy usage. However, if RIS is installed on the walls of a building that is close to a BS, the elements can be powered by WPT from the BS and RIS can use more elements for aiding the network. Likewise, a satellite-mounted RIS can rely on the sustainable solar power for its operation. Optimization of the number of active RIS elements considering the network requirements, available energy at the RIS, and the dynamics of energy harvesting is a challenging problem. 

\subsection{Impairments and Practical Limitations}
Hardware impairments (HIs), modeling imperfections, and other practical limitations pose great challenges on the design of RIS-enhanced INTENT networks. For example, it is common to assume continuous and perfect phase shift design for RIS elements; however, high resolution phase shifters are costly and employing them may compromise the cost-efficiency of the RIS. Therefore, in practice, only a limited phase shifts are available to choose from. Additionally, due to the complicated nature of optimization problems in RIS-aided systems, global optimal solutions are hardly achieved. Thus, the obtained phase shifts may differ from the actual optimum values. These two factors lead to residual phase noise at RIS elements which seriously degrades the optimization performance \cite{Xing}. Other limitations include HIs in RF components of the transceivers, frequent link interruptions in maritime communications, mutual coupling between RIS elements, random shadowing in satellite communications, etc. 

\subsection{Real-Life Experimentations}
To date, RIS-aided systems have been mainly evaluated using numerical simulations, where researchers develop theoretical frameworks for assessment of their proposed algorithms. The performance of RIS-enhanced systems in reality, however, might be ways different from what researchers observe in their MATLAB simulations. Therefore, it is safe to say that the numerical results provided in extensive research works on RIS cannot be trusted unless they are verified through real-life experiments. Nevertheless, carrying out testbed experiments of RIS-enhanced systems is far from straightforward, especially when it comes to non-terrestrial networks.

\section{Conclusion}
RIS is an ingenious technology which will break new grounds for the accomplishment of 6G-enabled INTENT networks and take wireless communications to the levels unthinkable today. This article presented the architecture of the forthcoming RIS-empowered INTENT networks and explicated the direct and indirect roles of RIS for furthering the performance of these networks. Simulation results were provided for evaluating the performance of RIS-enhanced NOMA-based satellite communications. We also presented the key challenges which can hinder the successful realization of 6G-enabled RIS-empowered systems in terrestrial and non-terrestrial environments. For future investigations in this promising area, researchers are encouraged to have a closer look into the challenges provided in this article and develop robust solutions to tackle these challenges.

\begin{IEEEbiography}{Parisa Ramezani}
received her B.S. degree in electrical engineering
from the Sharif University of Technology, Tehran, Iran, in 2014, and the M.Phil. and Ph.D. degrees from the University of Sydney, Sydney, NSW, Australia, in 2017 and 2022, respectively. She is currently a Postdoctoral researcher at KTH Royal Institute of Technology, Stockholm, Sweden. Her research interests include next-generation wireless networks, ultra massive MIMO communications, and reconfigurable intelligent surface-assisted communications. 
\end{IEEEbiography}

\begin{IEEEbiography}{Bin Lyu}
received the B.Eng. and Ph.D. degrees from Nanjing University of Posts and Telecommunications (NJUPT), Nanjing, China, in 2013 and 2018, respectively, where he is currently an Associate Professor. His research interests include Internet of Things, wireless powered communications, backscatter communications, reconfigurable intelligent surface, and mobile edge computing.
\end{IEEEbiography}

\begin{IEEEbiography}{Abbas Jamalipour}
received his Ph.D. degree in electrical engineering from Nagoya University, Japan, in 1996. He is a professor of ubiquitous mobile networking with the University of Sydney, Sydney, New South Wales, 2006, Australia. Since January 2022, he has been the editor-in-chief of IEEE Transactions on Vehicular Technology. He has authored nine technical books, 11 book chapters, and more than 550 technical papers and holds five patents, all in the area of wireless communications.
\end{IEEEbiography}

\end{document}